# A flexible approach for fat-water separation with bipolar readouts and correction of gradient-induced phase and amplitude effects.

Jorge Campos Pazmiño[1,2], Renée-Claude Bider[1], Véronique Fortier[1,3,4], and Ives R. Levesque[1,2,4,5]

[1]*Medical Physics Unit, McGill University, Montréal, QC, Canada*

[2]*Department of Physics, McGill University, Montréal, QC, Canada*

[3]*Department of Medical Imaging, McGill University Health Centre and Department of Radiology, McGill University, Montréal, QC, Canada*

[4]*Gerald Bronfman Department of Oncology, McGill University, Montréal, QC, Canada*

[5]*Research Institute of the McGill University Health Centre, Montréal, QC, Canada*

May 9, 2025

* Correspondence to:

Ives R. Levesque, PhD

Medical Physics Unit, McGill University

Cedars Cancer Centre - Glen Site, DS1.9326

1001 boul. Décarie

Montréal, QC H4A 3J1

ives.levesque@mcgill.ca

514-934-1934 ext. 48105

**Keywords:** Fat-water separation, bipolar echoes, chemical shift

**Abstract**

**Purpose:** To develop a fat-water separation approach that corrects bipolar readout gradient induced effects, without additional scans, that is compatible with any fat-water separation method.

**Theory and Methods:** The proposed approach combines joint fat-water separation of the odd and even echoes of a bipolar multi-echo gradient echo acquisition with an inverse problem to find least-squares estimates for phase and amplitude corrections to eliminate bipolar-induced effects. Optimization of sequence parameter selection through the calculation of the number of signal averages (NSA) with Cramér-Rao Bound theory (CRB) is presented. The application of the proposed approach is demonstrated with a graph cut optimization and further characterization of the accuracy was performed via Monte Carlo Simulations (MC). The proposed approach was tested in phantoms and in vivo. Proton density fat fraction maps (PDFF) were evaluated to quantify performance.

**Results:** NSA calculations suggest short TE$_1$ and ΔTE=1.5 ms as optimal alternatives for fat-water separation. MC simulations demonstrated accurate estimation of fat and water complex signals, $\psi$, and $R_2^*$ with mean relative error within 1%. In phantoms and in vivo, the proposed approach effectively eliminated effects induced by bipolar readout gradients, improving the outcome of the fat-water separation.

**Conclusion:** We proposed an approach to correct bipolar readout-induced effects that are detrimental for fat-water separation. This approach can extend the use of existing fat-water separation techniques designed for data acquired using unipolar readout gradients to data collected with bipolar readout gradients.

**Introduction**

Chemical shift encoded (CSE) fat-water separation in magnetic resonance imaging (MRI) exploits the difference in proton resonant frequency between fat and water molecules to generate fat- and water-specific signals through postprocessing. Data for CSE fat-water separation is commonly acquired with multi-echo sequences using readout gradient pulses with the same polarity, or "unipolar" readouts[1–3]. These require "flyback" gradient pulses to rewind the spatial encoding between each echo readout.

Bipolar readouts alternate the polarity of the readout gradient pulses and eliminate the need for flyback gradients. This enables shorter echo spacing, subsequently leading to unambiguous calculation of main magnetic field inhomogeneities[1,4–7]. Shorter echo spacing also facilitates imaging at high magnetic fields where the chemical shift-induced dephasing is faster[8]. Shorter echo spacing can also be used to acquire the required echoes within a short repetition time (TR), leading to short scan time.

Despite the benefits of using bipolar readout gradients, their use introduces challenges for fat-water separation not encountered with unipolar readouts. First, system non-idealities like gradient delays and eddy currents introduce phase errors that accrue in spatially opposite directions when alternating the readout gradient polarity[9]. This effect disrupts the phase difference between successive echoes (k-space echo misalignment) hampering fat-water separation. In contrast, unipolar readouts add a constant phase contribution to all echoes, avoiding this issue[4,9,10]. Second, the asymmetric non-flat frequency response of the receiver chain induces amplitude modulation in the MR signal, producing artifacts when readout polarity switches.[9,10]. Third, the reversal of polarity in bipolar readout gradient pulses causes field inhomogeneity-induced misregistration to occur in successive echoes with opposite readout direction[9]. Fourth, chemical shift-induced misregistration is also modulated in opposite readout directions due to the switch of polarity[9]. Polarity-induced field inhomogeneity and chemical shift effects produce signal-shifting, blurring, and distortion artifacts[4,9,11].

Various approaches exist for fat-water separation with bipolar readouts, each with specific limitations when correcting phase and amplitude effects. One approach corrects both first- and higher-order phase errors and amplitude modulation but requires an additional reference scan[10].

Others correct only first-order phase errors, and may be susceptible to noise—especially around fat fraction=0.5—when combining magnitude and complex-based fat-water separation[9,12,13]. Finally, techniques based on iterative decomposition of water and fat with echo asymmetry and least-squares estimation (IDEAL) can estimate both phase and amplitude errors but often depend heavily on initialization and lack spatial regularization[4,14]. An optimal technique would correct first- and high-order phase errors and amplitude modulation, without any additional scans and allow flexible selection of the fat-water separation method.

We present a novel approach for fat-water separation that corrects phase errors (first- and higher-order) and amplitude modulation induced by bipolar readout gradient pulses. First, we derived the theoretical framework to correct bipolar-induced effects, based on the joint postprocessing of odd and even echoes. Second, we optimized the echo timing for fat-water separation with odd and even echoes using a Cramér-Rao lower bound (CRB)-based formalism[15]. Subsequently, we assessed the precision and accuracy of the proposed bipolar fat-water separation approach using Monte Carlo (MC) numerical simulations. Finally, we tested the method experimentally in phantoms and in vivo to show accurate fat-water separation with bipolar readout gradient pulses and compared its performance to a unipolar approach.

**Theory**

Theory is presented for an MR acquisition using bipolar readout gradient pulses and $N$ echoes, under three assumptions.

1) Off-resonance frequency effects due to main field inhomogeneities $\psi(x, y)$ vary smoothly such that $\psi(x, y) \approx \psi(x \pm (\Delta x_{\text{fm}} + \Delta x_{\text{cs}}), y)$[9], where $\Delta x_{\text{fm}}$ and $\Delta x_{\text{cs}}$ are the spatial shifts in the readout direction due to local field inhomogeneities and chemical shift, respectively.

2) High readout bandwidth is used such that the spatial shifts $\Delta x_{\text{fm}}$ and $\Delta x_{\text{cs}}$ between echoes can be neglected.

3) A common $R_2^*$ term is used for fat and water relaxation, reducing the number of fit parameters and prioritizing the stability and noise performance in fat-water separation[16–19].

The complex signal in a reconstructed image voxel $S_n(x,y)$, for the $n^{\text{th}}$ echo at time $\text{TE}_n$, is given by:

$$S_n(x,y) = \left(W(x,y) + F(x,y) \sum_{m=1}^{M} \alpha_m e^{i2\pi \Delta f_m \text{TE}_n}\right) e^{i2\pi \psi(x,y)\text{TE}_n} e^{-R_2^*(x,y)\text{TE}_n} e^{(-1)^n i\theta(x,y)} \quad 1$$

where $W(x,y) = |W(x,y)|e^{i\phi_W(x,y)}$ and $F(x,y) = |F(x,y)|e^{i\phi_F(x,y)}$ are the complex water and fat signal components. The initial phases $\phi_W(x,y)$ and $\phi_F(x,y)$ account for the phase accrued during radio-frequency (RF) excitation and spoiling[20,21]. The sum $\sum_{m=1}^{M} \alpha_m e^{i2\pi \Delta f_m \text{TE}_n}$ represents the fat spectrum with $M$ distinct resonances, with relative amplitudes $\alpha_m$ and chemical shift frequency $\Delta f_m$. $\psi(x,y)$ and $R_2^*(x,y)$ represent the main field inhomogeneity and transverse relaxation, respectively. $\theta(x,y) \equiv \phi - i\varepsilon$ is a complex term containing the phase error $\phi$ and amplitude modulation $\varepsilon$ introduced by the bipolar readout [4,9,10,22].

To estimate $\phi$ and $\varepsilon$, we first separate the multi-echo data with opposite readout polarities, i.e. odd- and even-echo datasets. We then perform fat-water separation on the odd- and even-echo datasets using shared constraints on the $\psi(x,y)$ and $R_2^*(x,y)$ estimation. Subsequently, we use the complex fat and water signals from odd- and even-echo datasets to estimate the terms $\phi$ and $\varepsilon$ in the least-squares sense via a matrix equation. Finally, $\phi$ and $\varepsilon$ are used to correct the bipolar data to resemble unipolar data, following which any fat-water separation technique can be applied and find the unknown parameters of equation 1 ($|W|$, $|F|$, $\phi_W$, $\phi_F$, $\psi$, and $R_2^*$). The proposed method is illustrated in Figure 1.

Solving the fat-water separation for odd- and even-echo datasets with shared constraints, uses the system of equations:

$$S(\text{TE}_{\text{odd},n}) = \left(W_{\text{odd}} + F_{\text{odd}} \sum_{m=1}^{M} \alpha_m e^{i2\pi \Delta f_m \text{TE}_{\text{odd},n}}\right) e^{i2\pi \widehat{\psi} \text{TE}_{\text{odd},n}}$$
$$\quad 2$$
$$S(\text{TE}_{\text{even},n}) = \left(W_{\text{even}} + F_{\text{even}} \sum_{m=1}^{M} \alpha_m e^{i2\pi \Delta f_m \text{TE}_{\text{even},n}}\right) e^{i2\pi \widehat{\psi} \text{TE}_{\text{even},n}}$$

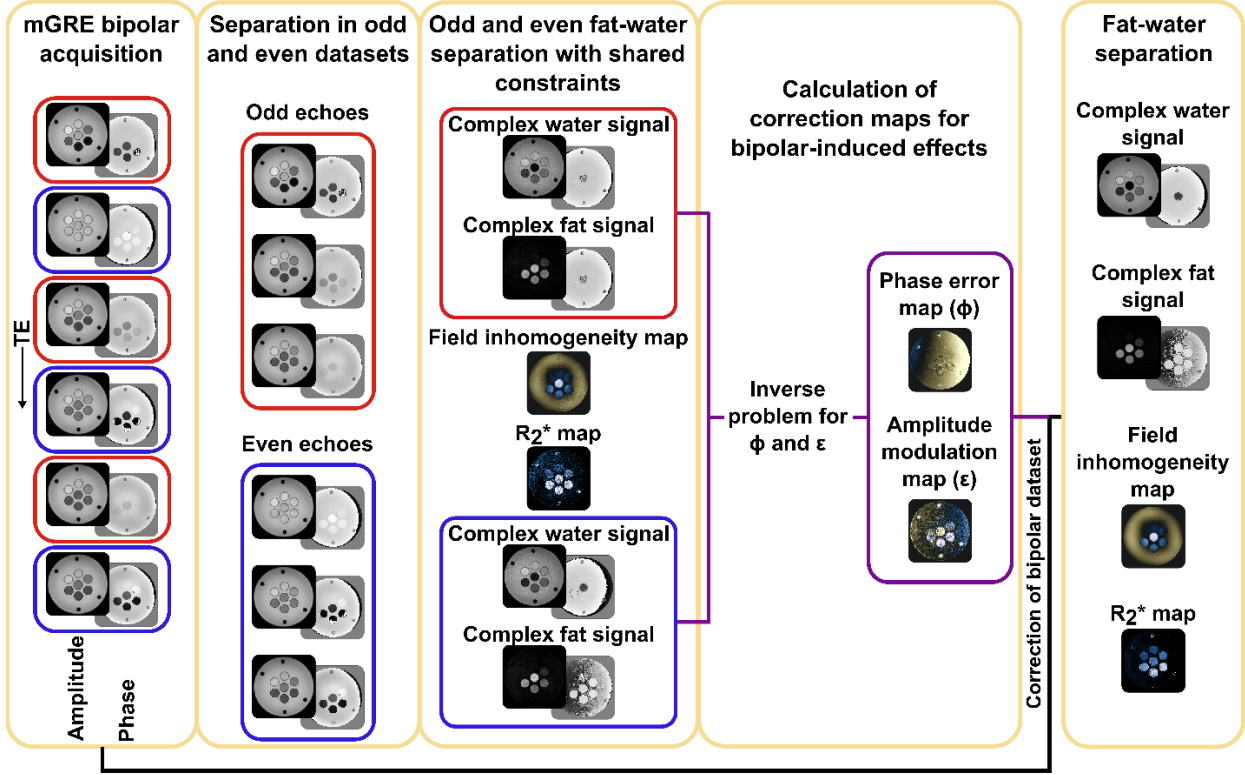

Figure 1: Schematic representation of fat-water separation technique for bipolar readouts. First, data is separated into odd- and even-echo datasets. Then, fat-water separation is performed jointly in these two sets and results are used to solve the inverse problem to estimate φ and ε maps. Finally, these maps are used to correct bipolar-induced effects and fat-water separation is applied to the synthetic unipolar dataset.

Equations 2 represents the data from odd- and even-echo datasets separately with subscripts. We have defined $W_{\text{odd}} \triangleq W(x,y)e^{-i\theta(x,y)}$, $F_{\text{odd}} \triangleq F(x,y)e^{-i\theta(x,y)}$, $W_{\text{even}} \triangleq W(x,y)e^{+i\theta(x,y)}$, and $F_{\text{even}} \triangleq F(x,y)e^{+i\theta(x,y)}$. The notation was simplified by defining $\hat{\psi} \triangleq \psi + i\frac{R_2^*}{2\pi}$, which contains field inhomogeneities inherent to the MRI system and from its shim settings for the imaging session, heterogeneity within the imaging subject, and transverse relaxation in the material ($R_2^*$). We can enforce the constraint $\hat{\psi}_{\text{odd}} = \hat{\psi}_{\text{even}} = \hat{\psi}$ as these are common to all echoes (odd or even). To jointly estimate parameters from Equation 2, it is represented as a matrix equation:

$$S = DA\rho \qquad 3$$

where:

$$S_{N\times 1} = \begin{bmatrix} S(\text{TE}_{\text{odd},1}) \\ S(\text{TE}_{\text{odd},3}) \\ \vdots \\ S(\text{TE}_{\text{even},2}) \\ S(\text{TE}_{\text{even},4}) \end{bmatrix} \qquad 4$$

$$D_{N\times N} = \begin{bmatrix} e^{i2\pi\widehat{\psi}\text{TE}_{\text{odd},1}} & 0 & \cdots & 0 & 0 & 0 \\ 0 & e^{i2\pi\widehat{\psi}\text{TE}_{\text{odd},3}} & \cdots & 0 & 0 & 0 \\ \vdots & \vdots & \ddots & \vdots & \vdots & \vdots \\ 0 & 0 & \cdots & e^{i2\pi\widehat{\psi}\text{TE}_{\text{even},2}} & 0 & 0 \\ 0 & 0 & \cdots & 0 & e^{i2\pi\widehat{\psi}\text{TE}_{\text{even},4}} & 0 \\ \vdots & \vdots & \vdots & \vdots & \vdots & \ddots \end{bmatrix} \qquad 5$$

$$A_{N\times 4} = \begin{bmatrix} 1 & \sum_{m=1}^{M} \alpha_m e^{i2\pi\Delta f_m \text{TE}_{\text{odd},1}} & 0 & 0 \\ 1 & \sum_{m=1}^{M} \alpha_m e^{i2\pi\Delta f_m \text{TE}_{\text{odd},3}} & 0 & 0 \\ \vdots & \vdots & 0 & 0 \\ 0 & 0 & 1 & \sum_{m=1}^{M} \alpha_m e^{i2\pi\Delta f_m \text{TE}_{\text{even},2}} \\ 0 & 0 & 1 & \sum_{m=1}^{M} \alpha_m e^{i2\pi\Delta f_m \text{TE}_{\text{even},4}} \\ 0 & 0 & \vdots & \vdots \end{bmatrix} \qquad 6$$

$$\boldsymbol{\rho}_{4\times 1} = \begin{bmatrix} W_{\text{odd}} \\ F_{\text{odd}} \\ W_{\text{even}} \\ F_{\text{even}} \end{bmatrix} \qquad 7$$

The inverse problem of estimating $\boldsymbol{\rho}$ from Equation 3 can be solved using various iterative optimization techniques commonly employed for fat-water separation with unipolar readout gradient pulses. This is possible since Equation 3 resembles the inverse problem of fat-water separation with unipolar readout gradients, with a modification to the dimensionality of the matrices in Equations 5 and 6. Changing the size of these two matrices does not affect the implementation of the optimization algorithm. As a proof of concept, we have solved the problem using a combination of variable projection (VARPRO) and graph-cut optimization methods [23,24]. The complex fat and water signals from odd- and even-echo datasets obtained using Equation 3 are then used in a matrix equation (Equation 8) that can be solved to estimate $\phi$ and $\varepsilon$.

$$\begin{bmatrix} i(1-\beta) & (1-\beta) \\ i\beta & \beta \end{bmatrix} \begin{bmatrix} \phi \\ \varepsilon \end{bmatrix} = \frac{1}{2} \begin{bmatrix} (1-\beta)[\ln(W_{\text{even}}) - \ln(W_{\text{odd}})] \\ (\beta)[\ln(F_{\text{even}}) - \ln(F_{\text{odd}})] \end{bmatrix} \qquad 8$$

The weight $\beta$ is selected as $\beta = 1$ for $\frac{|F_{\text{odd}}|}{|W_{\text{odd}}|+|F_{\text{odd}}|} \geq 0.5$ or $\beta = 0$ otherwise. The least-squares solution of Equation 8 was calculated with a constrained trust-region-reflective algorithm.

With $\phi$ and $\varepsilon$ known, the effect modelled by $\theta(x,y)$ in Equation 1 can be removed from the measured signal by multiplying by $e^{-(-1)^n i\theta(x,y)}$, generating a synthetic unipolar dataset. Finally, fat-water separation can be performed directly on this corrected dataset using any conventional method designed for unipolar readouts. For fat-water separation of unipolar data (measured and synthetic) in this work, we used a graph cut fat-water separation method[24].

**Methods**

**Performance analysis and optimization of fat-water separation with Cramér-Rao bounds**

CRB theory was used to calculate the number of signals averaged (NSA) for the fat-water separation step for odd- and even-echo datasets (Equation 3). The NSA quantifies the noise efficiency of the parameters estimated through fat-water separation[25]. The NSA is given by the ratio of a parameter's variance when it is the only free parameter in the model to its variance when all estimated parameters in the model are left free[15]. The numerator of the NSA is the reciprocal value of the corresponding diagonal entry in the Fisher information matrix (FIM)[26]. Meanwhile, the denominator is the corresponding diagonal entry of the inverse FIM[15,26]. We performed the calculations of the FIM utilizing theory from literature[15] modified to include the signal model in Equation 3 and extended to calculate the NSA for the magnitude and phase of the odd and even complex fat and water signals ($W_{\text{odd}}$, $W_{\text{even}}$, $F_{\text{odd}}$, $F_{\text{even}}$), $\psi$, and $R_2^*$. We performed the calculations with and without the constraint $\hat{\psi}_{\text{odd}} = \hat{\psi}_{\text{even}}$ to assess its effect on fat-water separation.

Optimal sequence parameters were selected by considering the NSA for each parameter across different proton density fat fractions (PDFF). Optimal sequence parameters were identified for the odd- and even-echo fat-water separation and then it was assumed that these sequence parameters are the best for the overall approach with bipolar readouts. NSAs were calculated for PDFF ∈ ]0,1[, first echo time $TE_1$ ∈ [0.5,2.5] ms, echo spacing ΔTE ∈ [0.5,2.5] ms, and number of echoes $N$= 6, 8, 10, 12, with TR≜$TE_1$+($N-1$)×ΔTE and flip angle FA=3 degrees. The fat-specific longitudinal relaxation rate $R_{1F}$=3.33 $s^{-1}$ was selected to be representative of bone marrow and subcutaneous fat at 3 T[27–29]. The water-specific longitudinal relaxation rate $R_{1W}$=0.67 $s^{-1}$ was selected to be representative of tissues like kidney, uterus, spleen, liver, and muscle at 3 T[28]. A six-resonance spectral model for peanut oil[12,30,31] was used. Parameters $\phi_W = \phi_F = \frac{\pi}{4}$, $R_2^*$=20 $s^{-1}$, $\psi$=40 Hz, $\phi$=0.02π, $\varepsilon$=0.03 were selected to match values used in literature[4,15].

To select the optimal sequence parameters, we first calculated the minimum NSA for each fat-water separation parameter across PDFF values, number of echoes, and $TE_1$–ΔTE pairs. For each $TE_1$ value, we then determined the ΔTE that maximized the minimum NSA. The shortest $TE_1$ was selected as optimal for each experiment since this boosted the minimum NSA. For each acquisition with $N$ echoes, the optimal ΔTE was considered as the mode of the ΔTE values that maximized the NSA across TE$_1$ values and parameters. For our experiments, we selected the mode across the $N$-echo acquisitions as a suitable choice for optimized ΔTE regardless of $N$. The search for optimal TE$_1$ was restricted to values between 1 and 2 ms, achievable in both phantom and in vivo experiments, while the ΔTE search was restricted between 1.1 and 2.2 ms due to a decrease in NSA and accuracy (characterized as explained in the next section) of the estimates outside this range.

**Accuracy and precision of the bipolar fat-water separation technique**

Monte Carlo (MC) simulations were used to evaluate the accuracy and precision of the fat-water separation estimates in the presence of noise across a range of phase errors $\phi \in$ [-π, π] radians and amplitude modulation $\varepsilon \in$ [-0.05,0.05]. The ranges of $\phi$ and $\varepsilon$ were selected based on literature[4]. TE$_1$=1 ms and ΔTE=1.5 ms were selected from NSA calculations (previous section). Six echoes were considered to present a worst-case scenario (minimum number of echoes required for the proposed technique). PDFF=0.50 and SNR=30 were used. All other parameters and assumptions were identical to the NSA calculations.

A complex signal was calculated using the model presented in Equation 1. White Gaussian noise was added to the real and imaginary parts of the signal and parameter estimation was repeated for 3000 noise realizations. The noise standard deviation was calculated as the ratio of the magnitude of the total signal calculated from Equation 1 for the first echo to the fixed SNR (=30). The relative error (= 100 × [estimated parameter – ground truth] / ground truth) and its standard deviation were used as metrics of accuracy and precision.

MC simulations were also used to verify that the sequence parameters, selected based on NSA calculations, led to accurate fat-water separation estimates. MC simulations were performed under the same conditions as in NSA calculations for PDFF = 0.1, 0.5, and 0.9. We analyzed the maximum absolute relative error across PDFF values and compared the sequence parameters that produced the most accurate estimates with those selected through NSA optimization.

**Phantom experiments**

The performance of the proposed fat-water separation method was evaluated with phantom experiments. A custom phantom was assembled inside a cylindrical enclosure (Cylindrical MagPhan, Phantom Laboratory, Greenwich, NY, USA) using 7 vials (50 mL centrifuge tubes, Corning®): 1 vial contained 3% agar by weight (MilliporeSigma Canada Ltd) mixed with gadolinium-based contrast agent (GBCA; gadobutrol, Gadovist 1.0 M, Bayer Healthcare), 5 vials contained fat-water emulsions, and 1 vial contained pure peanut oil. The emulsion was prepared following a published protocol[32] mixing deionized distilled water, 3% agar by weight, peanut oil (JVF Canada inc), GBCA, sodium benzoate (MilliporeSigma Canada Ltd), and non-ionic surfactants (Tween 20 and Span 80, MilliporeSigma Canada Ltd). The nominal fat volume fractions of the emulsions were 5%, 25%, 50%, 60%, and 75% and nominal GBCA concentrations were 0.19, 0.15, 0.10, 0.08, and 0.05 mM. The large phantom compartment was filled with a solution of GBCA (0.03 mM) and table salt (NaCl, 85 mM) in deionized distilled water. A diagram of the phantom is presented in Supplementary figure 1.

Phantom data was acquired using a 3 T MRI scanner (Ingenia, Philips Healthcare) with the vendor-supplied 15-channel head receive coil. All measurements were performed at room temperature. The temperature of the phantom was measured with a thermometer inserted in the large phantom compartment before (20°C) and after (21°C) MRI data acquisition. Data was acquired using a 3D multi-echo gradient echo (mGRE) sequence with RF and gradient spoiling. Acquisitions were performed separately using unipolar and bipolar readouts for two different experiments, described below.

The first experiment aimed to demonstrate accurate correction of bipolar-induced effects in fat-water separation. We scanned the phantom 4 times, using combinations of 6 or 10 echoes and unipolar or bipolar readout gradients. All datasets had matching FA= 3 degrees, $TE_1$=1.2 ms, $\Delta TE$=1.9 ms (minimum echo spacing for datasets with unipolar readouts), receiver bandwidth (rBW) = 1483.7 Hz/pixel ($\Delta x_{cs}$=0.3 pixels), voxel dimensions = 2×2×2 mm³, and a 3D field of view FOV = 406×406×104 mm³. The 6 and 10-echo datasets had TR=13 and 20 ms, respectively. No parallel imaging or signal averaging were used. Second-order shimming (pencil beam [PB]-

Volume, Philips) was applied in a volume covering the seven inserts in the phantom. PDFF maps were analyzed in cylindrical regions of interest (ROIs) within the phantom vials and an equally sized cylindrical ROI within the large phantom compartment, spanning five contiguous slices (ROI placement shown in Supplementary figure 1). Agreement between datasets was assessed using the intraclass correlation coefficient[33] (ICC; "poor"<0.5, "moderate"=0.5–0.75, "good"=0.75–0.9, "excellent">0.9). Accuracy of the proposed correction was evaluated in two comparisons: (1) unipolar datasets processed without corrections vs. the proposed approach (with correction), to assess any introduced bias; and (2) unipolar (without correction) vs. bipolar (with correction) datasets, to assess the effectiveness of the proposed approach.

The second experiment evaluated the performance of the proposed approach with different ΔTE. We compared fat-water separation obtained for data collected with either the optimal ΔTE (determined through NSAs) or with the minimum ΔTE achieved for bipolar readouts. We scanned the phantom twice, first with the optimal ΔTE=1.5 ms (TR=11 ms) and second with the minimum ΔTE=0.9 ms (TR=8 ms). All other parameters were unchanged from the first phantom experiment. PDFF maps were analyzed within the same ROIs (Supplementary figure 1). Precision was quantified using the interquartile range and agreement was assessed using Bland-Altman plots.

All datasets were postprocessed with a graph cut fat-water separation technique[24], without corrections for bipolar induced effects, and with the proposed technique. A six-resonance spectral model for peanut oil[12,30,31] was used. PDFF maps were calculated with a magnitude discrimination method[34] to mitigate potential noise-bias.

**In vivo experiments**

The proposed approach was evaluated in vivo in the knee of a healthy female volunteer (25 years old) and the abdomen of a healthy male volunteer (30 years old). Both participants provided informed consent. All scans were performed on the same 3 T scanner (Ingenia, Philips Healthcare) used in phantom experiments. The knee datasets were acquired using the vendor-supplied 8-channel knee receive coil. A lower rBW = 719.0 Hz/pixel ($\Delta x_{cs}$=0.6 pixels) was used to enable acquisitions with smaller voxels. The abdominal datasets were acquired using the vendor-supplied body array receive coils (anterior and posterior). We used a similar rBW = 1483.7 Hz/pixel ($\Delta x_{cs}$=0.3 pixels) as in phantom experiments. Other sequence parameters for all acquisitions are summarized in Table 1. For all in vivo datasets, we used a six-resonance spectral model for human adipose tissue[30,35]. Variation in results are expected depending on the selection of the fat spectrum[36] and this model was selected for all datasets since it produced results free of fat-water swaps in all the considered anatomies. Different anatomies were used to test the proposed approach in different scenarios and to demonstrate strategies for improving fat-water separation using bipolar readout gradient pulses.

Table 1: Sequence parameters for in vivo experiments. Table reports the 'reconstructed' voxel size, which was selected to match the acquisition voxel size as closely as possible. All datasets were collected with no parallel imaging and no signal averaging.

**Knee datasets**

| Readout | Echoes | FA [°] | TE$_1$ [ms] | ΔTE [ms] | TR [ms] | Voxel size [mm³] | FOV [mm³] | Scan time [min:sec] |
|---|---|---|---|---|---|---|---|---|
| Unipolar | 6 | 3 | 1.87 | 3.2 | 42 | 0.8×0.8×2.0 | 170×170×140 | 10:23 |
| Bipolar | 6 | 3 | 1.87 | 1.7 | 23 | 0.8×0.8×2.0 | 170×170×140 | 5:40 |

**Abdomen datasets**

| Readout | Echoes | FA [°] | TE$_1$ [ms] | ΔTE [ms] | TR [ms] | Voxel size [mm³] | FOV [mm³] | Scan time [sec] |
|---|---|---|---|---|---|---|---|---|
| Unipolar | 6 | 3 | 1.1 | 1.5 | 11.0 | 2.3×2.3×5.0 | 300×206×106 | 31 |
| Bipolar | 6 | 3 | 1.1 | 0.9 | 7.2 | 2.3×2.3×5.0 | 300×206×106 | 18.5 |
| Bipolar | 10 | 3 | 1.1 | 0.9 | 11.0 | 2.3×2.3×5.0 | 300×206×106 | 28 |

For the knee datasets, bipolar readout gradients enabled a shorter ΔTE, which should improve fat-water separation, reduce TR, and decrease scan time. The knee was scanned twice—with unipolar and bipolar readout gradient pulses—with the same number of echoes (6) and the minimum ΔTE for each readout. Graph-cut fat-water separation (without correction) was applied to both the unipolar and bipolar datasets, while the approach with correction for bipolar gradient effects was applied only to the bipolar dataset. The root mean square error (RMSE) was used to evaluate the goodness of fit for each model, and PDFF difference maps were analyzed to assess the impact of bipolar gradient corrections.

For the abdominal datasets, the reduced ΔTE enabled by bipolar readouts was used in two ways: 1) to reduce TR and scan time while maintaining the number of echoes (6) or 2) to improve the quality of the fat-water separation by including more echoes within a given TR and scan time. Data were collected under a single breath-hold, with full coverage of the liver. Three datasets were collected: one with unipolar readouts and two with bipolar readouts. PDFFs within ROIs in the liver, spleen, and subcutaneous fat were compared using Bland-Altman plots. Circular ROIs were manually selected in the liver and spleen. For subcutaneous fat, an ROI was generated within a manually selected rectangular region by thresholding voxels with PDFF > 0.80. ROIs are depicted in Supplementary figure 2**Error! Reference source not found.**.

**Results**
**Performance analysis and optimization of fat-water separation with Cramér-Rao bounds**

Constraining the odd- and even-echo fat-water separation—setting $\hat{\psi}_{\text{odd}} = \hat{\psi}_{\text{even}}$—increased the NSA of the estimates compared to unconstrained postprocessing. Figure 2 shows the NSA calculated with and without constraints for the estimation of magnitude and phase of the fat and water signals, ψ, and $R_2^*$ map. The NSA for water and fat signals averaged across PDFF values increases by 64% and 46% for the odd and even datasets, respectively. For $\psi$ and $R_2^*$, the NSA increases by 5%.

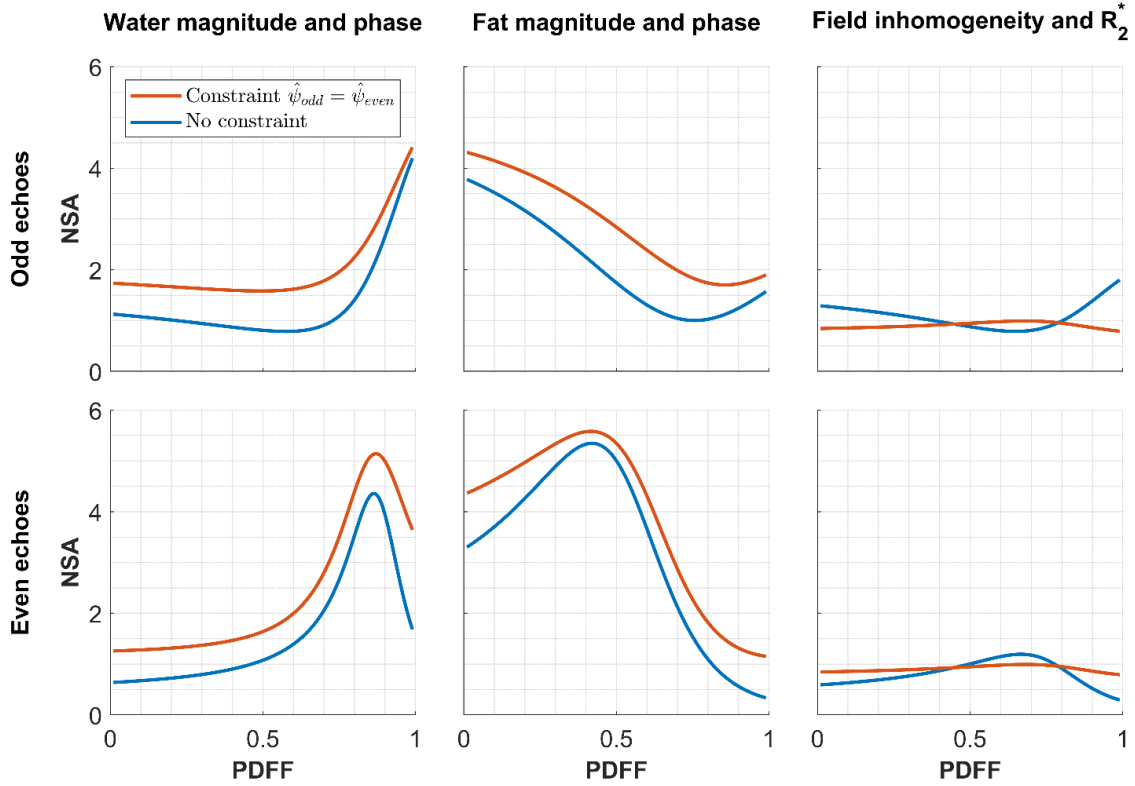

Figure 2: NSA comparison for fat-water separation estimates with odd- and even-echo datasets with and without constraints. Results for magnitude and phase of the water and fat signals, and for $\psi$ and $R_2^*$ are presented in single plots since the NSAs are the same for each case. Calculations done for 6 echoes, $TE_1$=1 ms and $\Delta TE$=1.5 ms. Orange line: NSA calculated for $\hat{\psi}_{odd} = \hat{\psi}_{even}$. Blue line: NSA calculated without constraints. The constraint increases the NSA of the fat and water signals estimates after fat-water separation.

NSA calculations suggest that the shortest $TE_1$ and $\Delta TE$=1.5 ms are optimal for odd- and even-echoes fat-water separation. Figure 3 presents heat maps for the minimum NSA (across PDFF), demonstrating that a short $TE_1$ enhances the minimum NSA. Based on this, we used $TE_1$=1 ms for all subsequent numerical simulations, as it is a reasonable $TE_1$ for phantom and in vivo experiments. According to our selection criteria, $\Delta TE$=1.7, 1.5, 1.5, and 1.4 ms are optimal for the 6, 8, 10, and 12-echo acquisitions (Figure 3). We selected $\Delta TE$=1.5 for all experiments since this offers a good NSA performance across all estimates and number of echoes. The maximum absolute relative error across PDFFs (heat maps in Supplementary figure 3) further supports the selection of $\Delta TE$=1.5 ms for 6, 8, 10, or 12-echo acquisitions, since it offers great accuracy for fat-water separation.

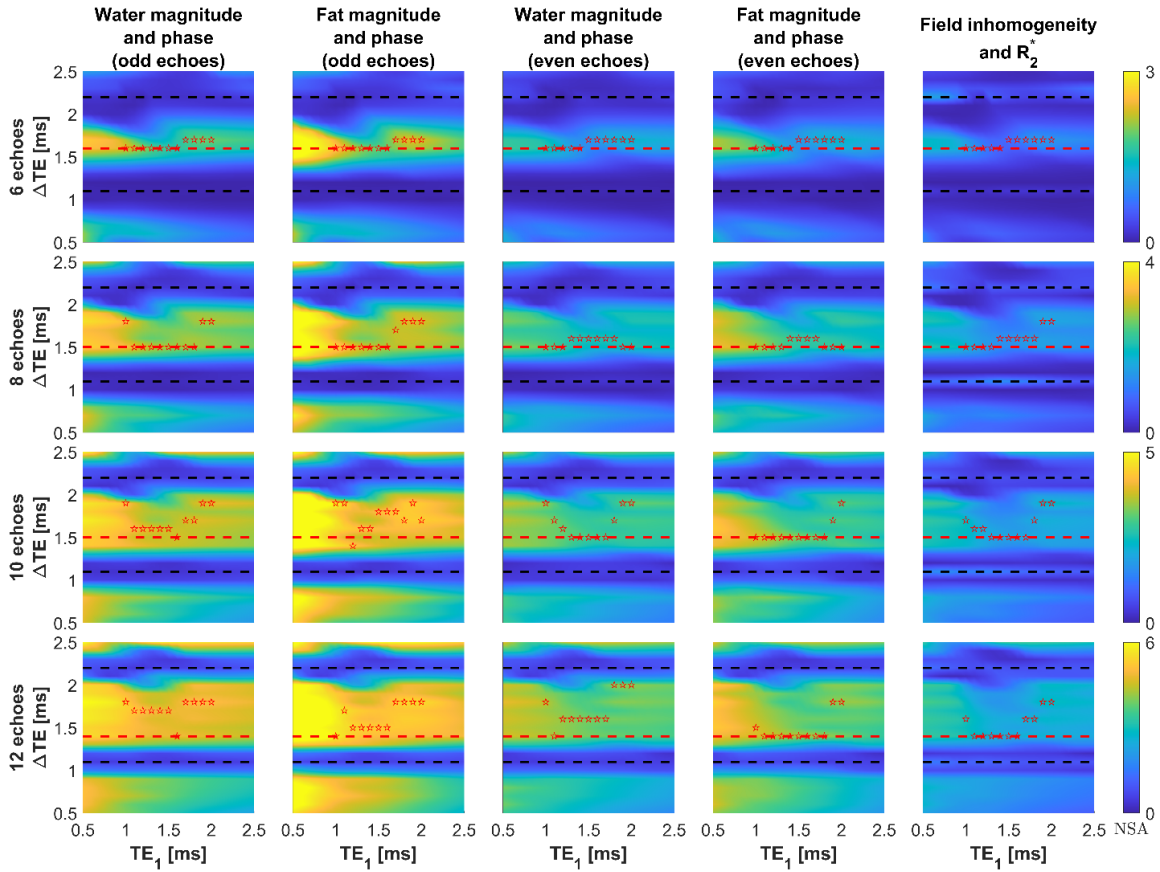

Figure 3: Minimum NSA (across PDFF) as function of TE$_1$ and ΔTE. Rows (top to bottom): Calculations for $N$=6, 8, 10, and 12 echoes. Columns: Fat-water separation estimates for odd- and even-echo datasets. Results for magnitude and phase of the water and fat signals, and for $\psi$ and $R_2^*$, are presented in a single plot since the NSAs are the same for each case. NSA values are displayed between 0 and $N/2$. Stars: Optimal ΔTE for each TE$_1$. Dashed lines: range for search of optimal ΔTE (black) and optimal ΔTE (red) for each $N$. ΔTE=1.5 ms is the mode across the acquisitions with different $N$ and is proposed as a suitable choice to optimize fat-water separation.

**Accuracy and precision of the bipolar fat-water separation technique**

In MC simulations, the proposed approach was accurate and precise in the presence of noise and for a wide range of $\phi$ and $\varepsilon$. As seen in **Error! Reference source not found.**, the mean relative difference for the magnitudes and phases of the fat and water signals, and for $\psi$ and $R_2^*$, remained <1% for $\phi \in$[-π, π] and $\varepsilon \in$ [-0.05,0.05], revealing high accuracy. Estimates of the fat and water signal magnitude and phase and $\psi$ were also precise, with standard deviation of the relative error within ±1%. $R_2^*$ was the least precise with a standard deviation of the relative error of ±7%.

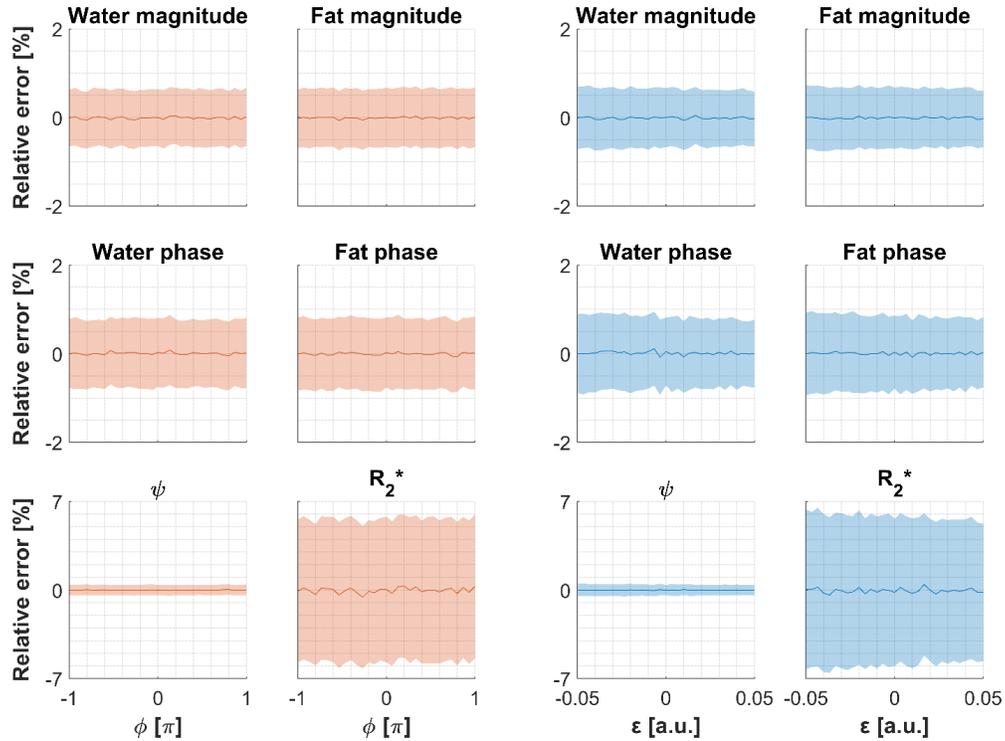

Figure 4: Monte Carlo simulation for the proposed bipolar fat-water separation. Left two columns (orange): Results for $\phi \in [-\pi, \pi]$ and $\varepsilon$ =0.03. Right two columns (blue): Results for $\phi$ = 0.02 $\pi$ and $\varepsilon \in [-0.05, 0.05]$. Solid line: mean relative difference. Shaded region: Standard deviation of the relative difference. Simulation results indicate that the complex fat and water signals and $\psi$ estimates are precise and accurate across a realistic range of $\phi$ and $\varepsilon$ values.

**Phantom experiments**

The proposed approach enables accurate fat-water separation and eliminates phase errors and amplitude modulation induced by bipolar readout gradient pulses. Maps in Figure 5 a) show that, for a 6-echo dataset with unipolar readouts, the fat-water separation without corrections and the proposed approach (with corrections) are equivalent. This is supported by the data plotted in Figure 5 b) where the ICC>0.9 shows excellent agreement. Furthermore, maps in Figure 5 c) show that standard fat-water separation (without correction) on bipolar data results in spurious fat signal within the large water compartment. The proposed approach effectively eliminates these artifacts, as evidenced in Figure 5 c) and plot d), where ICC > 0.9 shows excellent agreement between the PDFF maps from the unipolar dataset (processed with GC without corrections) and the bipolar dataset (processed with corrections). Findings for the 10-echo dataset were consistent with those observed for the 6-echo data (Supplementary figure 4).

$\phi$ and $\varepsilon$ maps used to correct bipolar-induced effects exhibit a distinct spatial dependence, primarily along the readout direction. As seen in Figure 6, the $\phi$ and $\varepsilon$ maps obtained for the 6-echo unipolar acquisition (same data as in Figure 5) remain close to zero across most of the phantom, except in regions with a high fat fraction (nominal fat fraction ≥ 0.5) similar to a prior report[4]. In contrast, $\phi$ and $\varepsilon$ maps for the bipolar dataset display a gradient along the readout direction, also consistent with findings from other studies[4,10]. These $\phi$ and $\varepsilon$ maps are derived from the complex fat and water signals of the odd and even echoes, which can be viewed in Supplementary figure 5, along with the synthetic unipolar dataset from bipolar acquisitions. Supplementary figure 5 highlights a noticeable reduction in noise and artifacts in the fat-water separation when using the synthetic unipolar dataset, demonstrating the effectiveness of the correction method.

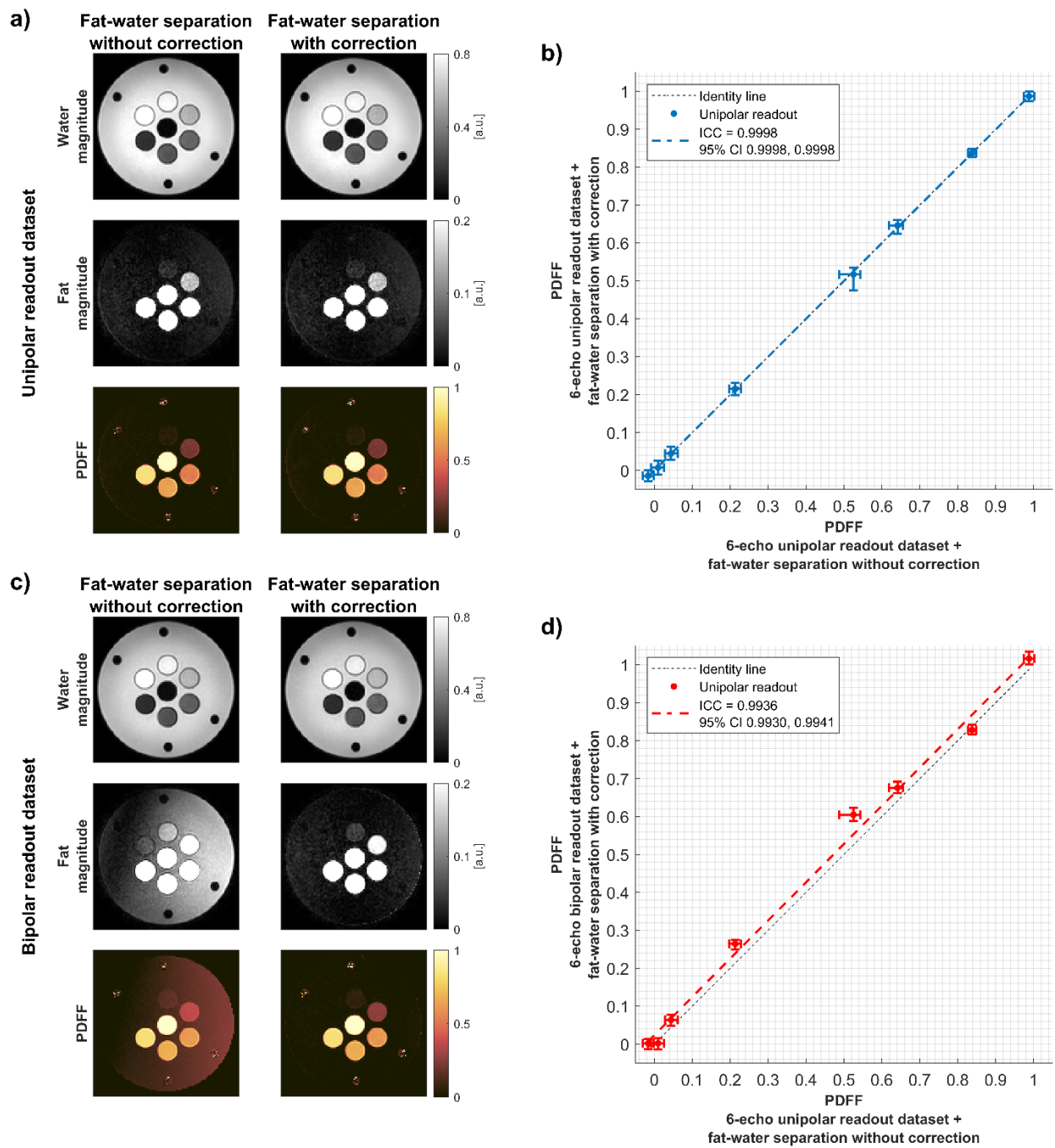

Figure 5: Fat-water separation performance in 6-echo data acquired with unipolar and bipolar readouts. Data processing was performed with the graph-cut fat-water separation without corrections for bipolar-induced effects and with the proposed approach (with corrections). a) Maps for the unipolar readout dataset processed without and with corrections, resulting in equivalent maps. b) PDFF comparison for the unipolar dataset processed with both techniques. c) Maps for the bipolar readout dataset, highlighting the impact of gradient effect corrections. d) PDFF comparison for unipolar and bipolar datasets, without and with corrections, respectively. In b) and d), markers indicate the median PDFF and error bars represent the 25th–75th percentiles.

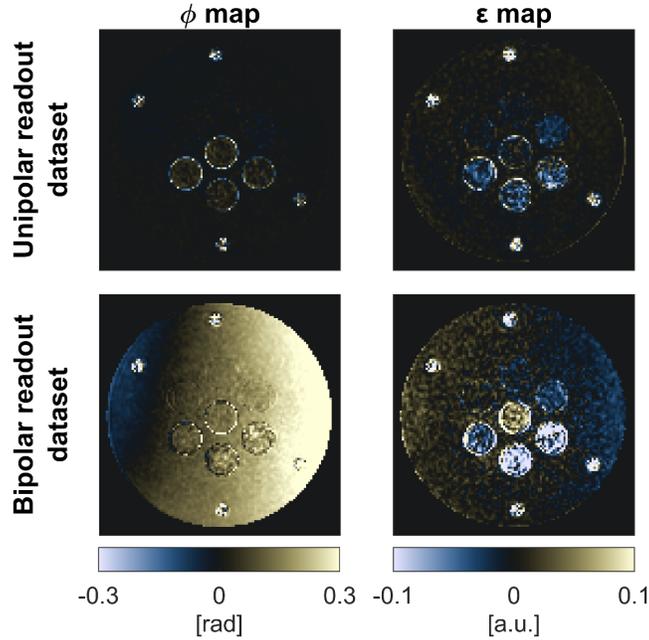

Figure 6: Estimated phase errors ($\phi$) and amplitude modulation ($\varepsilon$) maps used to correct bipolar-induced effects. Top row: Correction maps for unipolar readout dataset with 6 echoes. Bottom row: Correction maps for bipolar readout dataset with 6 echoes. Primarily, $\phi$ and $\varepsilon$ maps change linearly in the readout direction (left to right). Two different datasets are presented to illustrate the variability of the correction maps across datasets.

Echo spacing optimization using NSA calculations leads to improved precision in fat-water separation with bipolar readouts and the proposed approach. Fat-water separation results for 6- and 10-echo bipolar readout datasets and two different echo spacings can be compared in Figure 7. As shown in Figure 7 a), using the minimum ΔTE achievable for this experiment (0.9 ms) increased the noise in both fat and water magnitude images, compared to images from the optimized ΔTE (1.5 ms). This observation is supported by Figure 7 b), which displays a broader distribution of PDFF values for the 6-echo datasets acquired with the minimal echo spacing. The inter-quartile range (IQR) of PDFF for the 6-echo datasets, averaged across all ROIs, was 43% lower for the results with optimal ΔTE. For the 10-echo datasets, the IQR decreased by an average of 8% when using the optimal ΔTE. Figure 7 b) also shows variations in the average PDFF that depend on echo number and spacing. Bland-Altman plots comparing the PDFF from 6-echo and 10-echo datasets with different echo spacing (Supplementary figures 6 and 7) show that the largest mean difference across all inserts is 0.022.

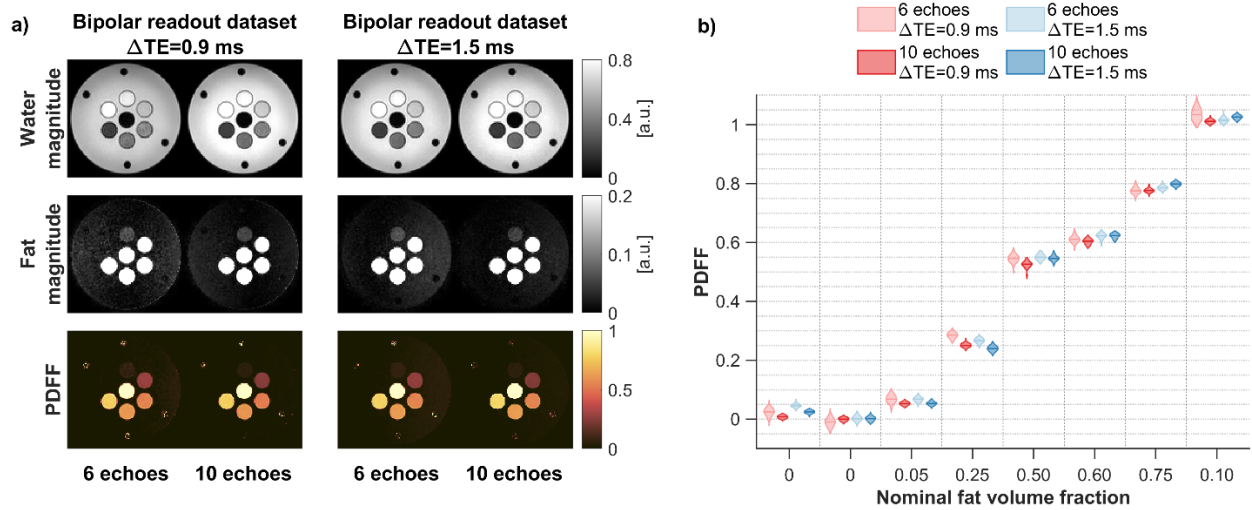

Figure 7: Comparison of fat-water separation results depending on echo spacing. a) Fat-water separation results for 6- and 10-echo datasets collected with bipolar readout gradients and ΔTE=0.9 ms (minimum ΔTE for the experiment) and ΔTE=1.5 ms (optimal ΔTE). b) Violin plots comparing PDFF maps evaluated in ROIs corresponding to a portion of the large phantom compartment and the 7 inserts.

**In vivo experiments**

In vivo, the proposed approach enabled the use of bipolar readouts for knee imaging by effectively correcting phase errors and amplitude modulation effects that typically degrade fat-water separation when using bipolar readout gradients. Figure 8 a) shows that the proposed approach successfully removed the spurious fat signal observed in muscle regions in the uncorrected fat-water separation results. The corresponding RMSE maps indicate that correcting these errors provided a better fit to the measured data. Errors in PDFF present a spatial dependence that stems from the spatial distribution of $\phi$ show in Figure 8 b). The PDFF difference maps comparing fat-water separation results for unipolar and bipolar datasets are shown in Figure 8 c).

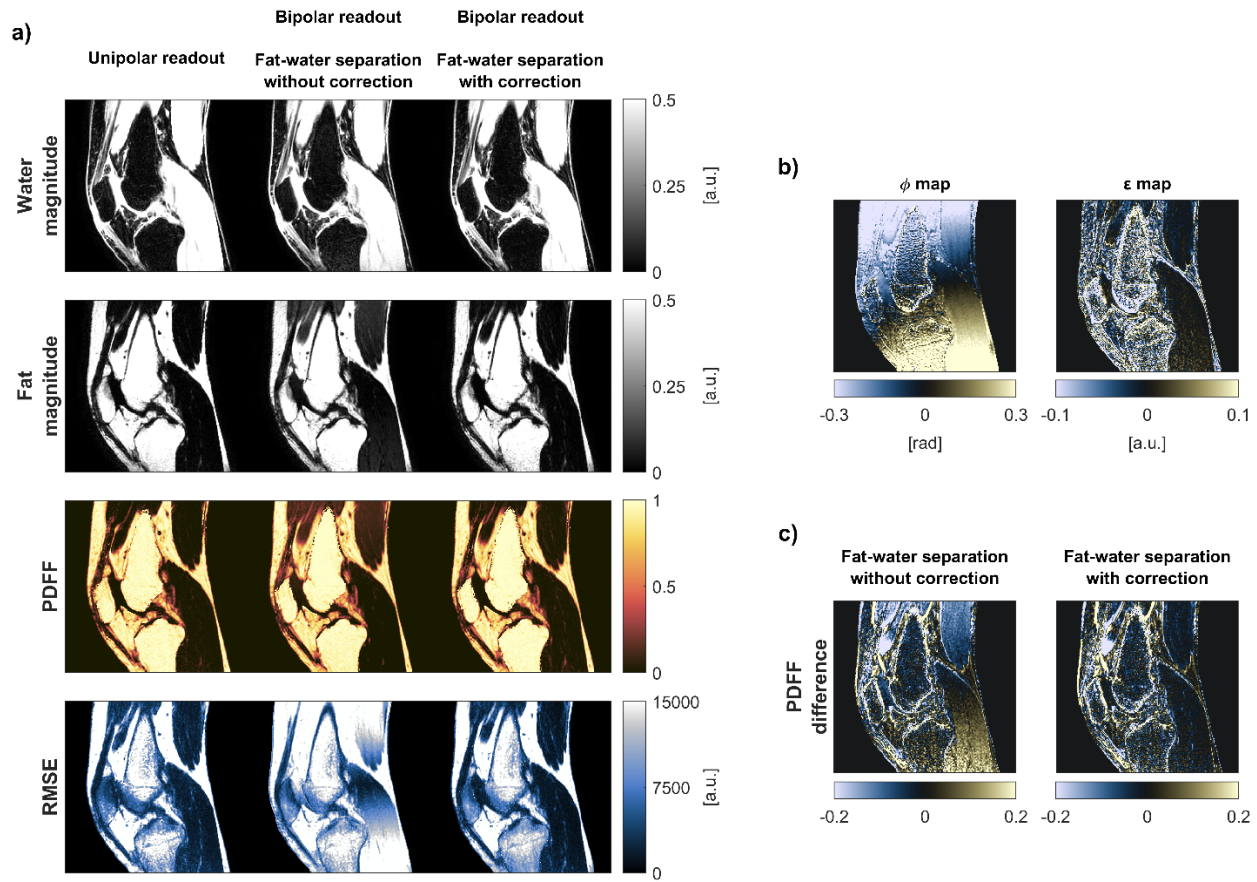

Figure 8: Fat-water separation results for knee datasets collected with unipolar and bipolar readout gradients. a) Water and fat magnitude images, PDFF, and RMSE maps for datasets postprocessed without and with corrections for bipolar-induced effects. b) Phase errors ($\phi$) and amplitude modulation ($\varepsilon$) maps for fat-water separation with corrections for the bipolar dataset. c) PDFF difference maps between results for unipolar and bipolar readout datasets. Left: Difference map with PDFF map without corrections. Right: PDFF map with corrections. The proposed approach eliminates spurious PDFF values.

The proposed approach effectively eliminated bipolar-induced artifacts in the abdominal images, while enabling either the reduction of scan time or the improvement of the fat-water separation. Figure 9 presents fat-water separation results for datasets collected using unipolar and bipolar readout gradients. For the bipolar dataset, results are shown both without and with corrections for bipolar-induced effects. Failing to correct for bipolar-induced effects leads to a spurious increase in the PDFF when compared to the unipolar dataset or the bipolar dataset with corrections. This effect is especially noticeable in the dataset with a reduced TR=7.2 ms (and shortest scan time). This artificial fat signal increase exhibited a spatial pattern that corresponded closely with the phase error ($\phi$ map not shown), consistent with the findings from the knee imaging experiments. Quantitative comparison of PDFF showed differences between the unipolar dataset and the 6-echo bipolar (TR=7.2 ms) dataset without corrections of 0.11 CI: [0.11, 0.12] , 0.19 CI: [0.18,0.20], and 0.11 CI: [0.09,0.13] in ROIs in the liver, spleen, and subcutaneous fat, respectively (Bland-Altman plots in Supplementary figure 8). When comparing PDFFs from the unipolar dataset and 6-echo bipolar dataset with corrections, the mean differences were lower in all three ROIs: 0.009 CI: [0.005, 0.013], -0.010 CI: [-0.019, -0.001], and 0.061 CI: [0.042, 0.081]. These improvements are supported by the lower RMSE (maps in Figure 9) that shows a better fit to the measured signal when using the proposed approach. When comparing the PDFFs from the unipolar 10-echo bipolar (TR=11 ms) datasets, the discrepancies between them also decreased when correcting for bipolar induced effects (Blant-Altman plots in Supplementary figure 9). Most notably, the differences evaluated in the liver and subcutaneous fat were smaller: 0.005 CI: [0.002,0.008], and 0.036 CI: [0.028,0.044], respectively. Moreover, the limits of agreement in the plots across all three ROIs also decreased. Taken together, these results reflect the improvement of the fat-water separation.

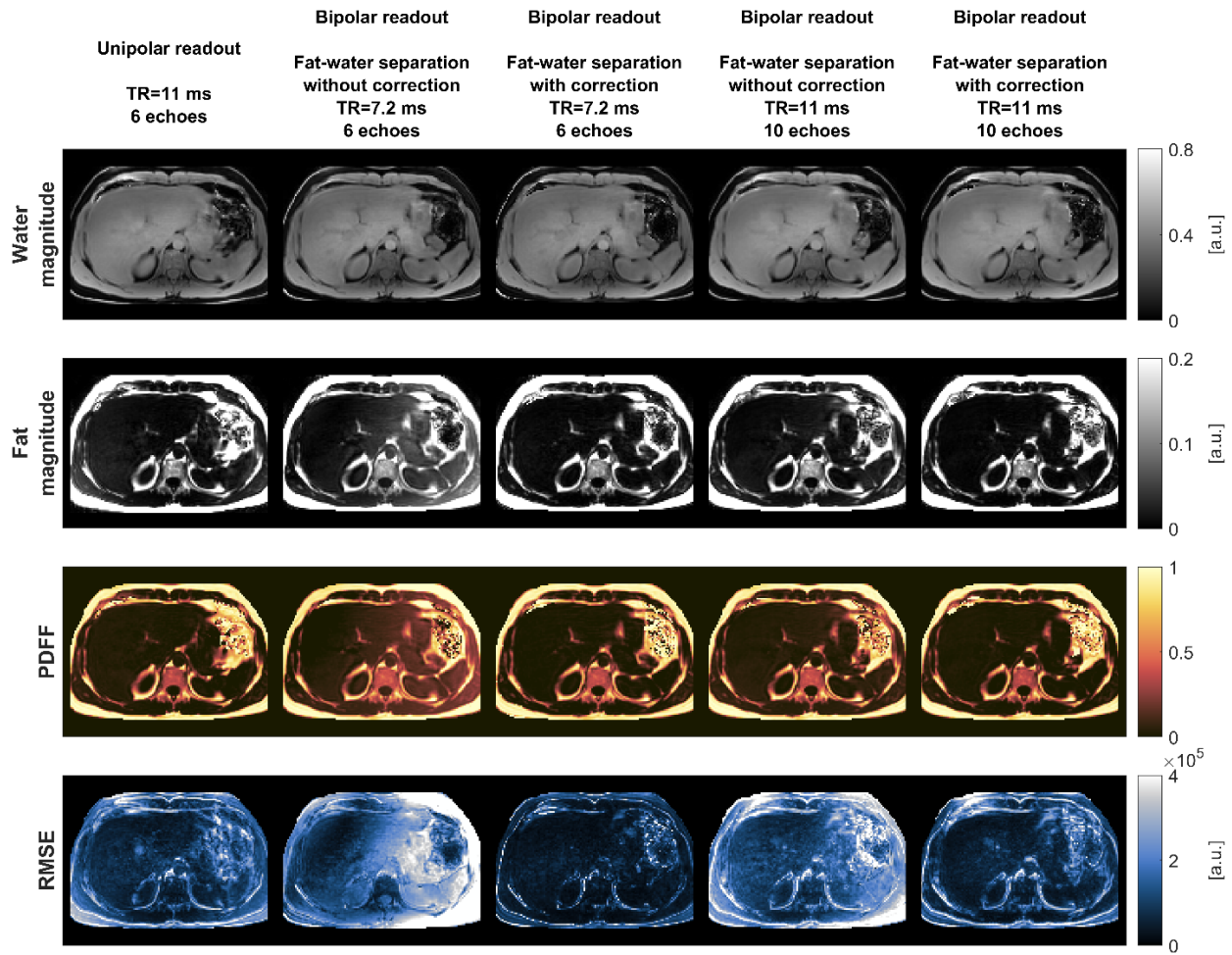

Figure 9: Fat-water separation results for abdomen datasets collected with unipolar and bipolar readout gradients, showing water and fat magnitude images, PDFF, and RMSE maps for datasets processed without and with corrections for bipolar-induced effects. Two bipolar datasets are included: 1) 6 echoes, TR=7.2 ms and 2) 10 echoes, TR=11 ms. These results illustrate how bipolar readout gradients can be used to reduce TR and thereby scan time or to maximize the number of echoes while keeping the TR fixed. The proposed approach eliminates artifacts in both scenarios.

**Discussion**

This work proposed a novel approach for fat-water separation on data acquired with bipolar readout gradient pulses that corrects disruptive phase errors and amplitude modulation effects[9]. This approach addresses the limitations of previously created techniques[4,9,10,13] in three ways. First, it works with data from a single acquisition with bipolar readouts gradient pulses without any extra acquisitions. Second, it corrects phase errors (first- and higher-order) and amplitude effects. Finally, it can be paired with any optimization technique for fat-water separation and is not limited to IDEAL-type approaches.

MC numerical simulations showed that the proposed approach is accurate in the presence of noise and bipolar-induced effects. Phantom and in vivo data showed better fat-water separation for datasets that used bipolar readout gradients, possibly due to the optimized echo spacing. The most notable improvements from the proposed approach were the reduction of artifacts induced by bipolar readouts and the elimination of spurious signal that led to artificially increased PDFF values. Notably, when combined with the graph-cut- fat-water separation approach[24], this method effectively prevented fat-water swaps commonly reported elsewhere[4]. Results for phantom and in vivo experiments with bipolar echoes were free of fat-water swaps. Finally, phantom experiments showed that the proposed approach is accurate and does not introduce bias in the estimation of PDFF, which suggests excellent correction for phase (first and higher order) and amplitude effects induced by bipolar readouts[4].

CRB-based calculations of NSA enabled the estimation of optimal echo timing parameters for the proposed approach with a variety of echo train lengths (6–12 echoes). Phantom experiments demonstrated that the most precise results were obtained from bipolar readouts with optimal echo spacing, when compared to the minimum echo spacing for unipolar or bipolar readouts. However, similar to results from Peterson and Manson[4], NSA heat maps presented bands in which NSA steeply decreased: these bands of reduced NSA limit the reduction of echo spacing despite the elimination of the flyback gradient, leading to optimal echo spacing larger than the minimum echo spacing. This ultimately poses a tradeoff between scan time reduction and optimal performance of the fat-water separation with bipolar readout gradients.

The proposed approach does not include strategies for removing field inhomogeneity and chemical shift-induced misregistration. These effects are inversely proportional to the readout bandwidth per pixel[9], such that we collected data with sufficiently high bandwidth to neglect them. Modifications could be implemented to enable the use of any readout bandwidth. For example, an image warping procedure to eliminate field inhomogeneity-induced misregistration could be incorporated[9,37]. Fat-water separation could be performed in k-space to eliminate chemical shift-induced misregistration between readouts[9,11,38].

The proposed approach enables fat-water separation for bipolar readout gradients using any state-of-the-art fat-water separation technique designed for unipolar readouts. The significance of this work is to provide a clear theoretical framework to extend techniques originally limited to unipolar readouts and to present experimental evidence of improved performance when using bipolar readout gradient schemes. This contribution can improve the versatility of fat-water separation techniques. However, we highlight that this approach will not correct for drawbacks inherent to the fat-water separation technique itself. In this paper, the graph-cut fat-water separation technique[24] was used because 1) code is widely available as part of the ISMRM fat-water separation toolbox[35] and 2) it is a widely used technique in recent literature[36]. Future work could consider other fat-water separation techniques in a systematic comparison to characterize their performance with bipolar readouts.

**Conclusion**

We proposed an approach to correct the measurement errors affecting the phase and amplitude of data from mGRE MR acquisitions with bipolar readout gradient pulse schemes, used for fat-water separated imaging. The proposed approach can extend the use of existing fat-water separation techniques designed for unipolar readout gradients. We presented a clear theoretical framework, sequence parameter optimization, and experimental evidence of the performance in fat-water separation when using the proposed approach.

**Acknowledgments**

The authors acknowledge the developers of the ISMRM fat-water toolbox (http://www.ismrm.org/workshops/FatWater12/data.htm) the developers of the 'Crameri' scientific colormaps (https://zenodo.org/records/5501399), Norma Ybarra for her contributions during phantom fabrication, and members of the MR Methods Research Group (McGill University) for useful discussion. Jorge Campos Pazmiño acknowledges funding provided from the Fonds de recherche du Québec – Nature et technologies (FRQNT; B2X - *Bourse de doctorat en recherche*). This project was funded by a Discovery Grant from the Natural Science and

**Data availability statement**

Source code to perform the bipolar gradient correction outlined in this work is shared on GitLab https://gitlab.com/MPUmri/bipolar_fat_water_separation.git. This code requires a working version of the ISMRM Fat-Water Separation Toolbox. Data from the phantom experiments are shared on the Open Science Foundation https://osf.io/bavk7/.

**Supplementary Information**

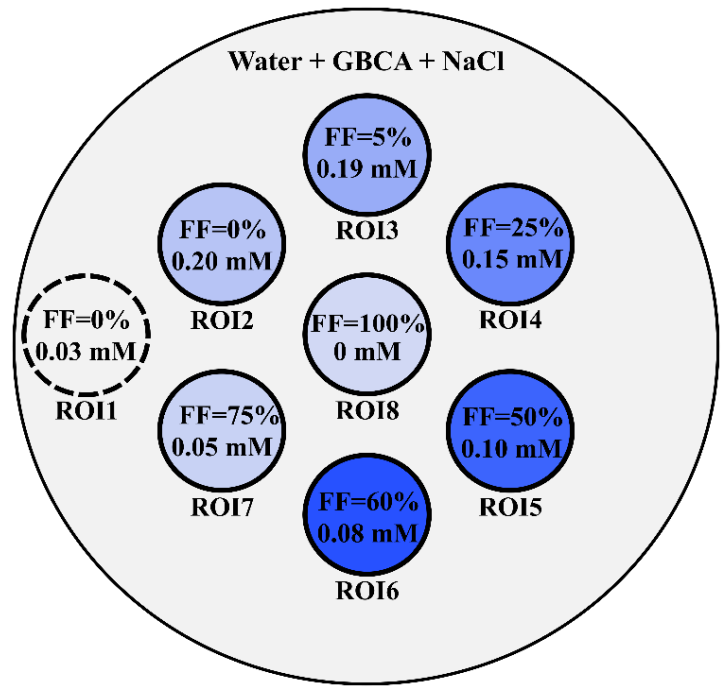

**Supplementary figure 1:** Phantom diagram. The phantom was assembled by placing seven 50 mL vials inside a large cylindrical enclosure. Regions of interest (ROIs) within the phantom were selected to cover a small portion of the large compartment (ROI1) and the vials (ROI2-ROI8). Diagram shows the nominal fat volume fractions and GBCA concentrations for all ROIs.

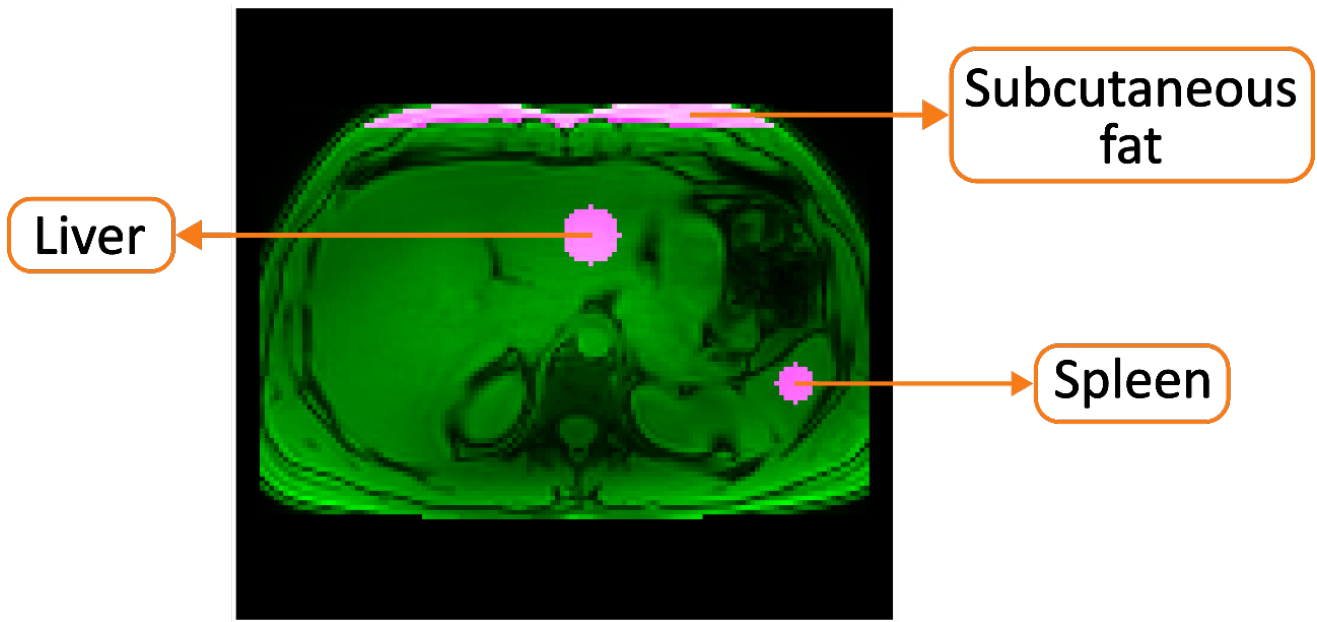

**Supplementary figure 2:** Liver, spleen, and subcutaneous fat ROIs for the quantitative analysis of PDFF maps for abdominal datasets.

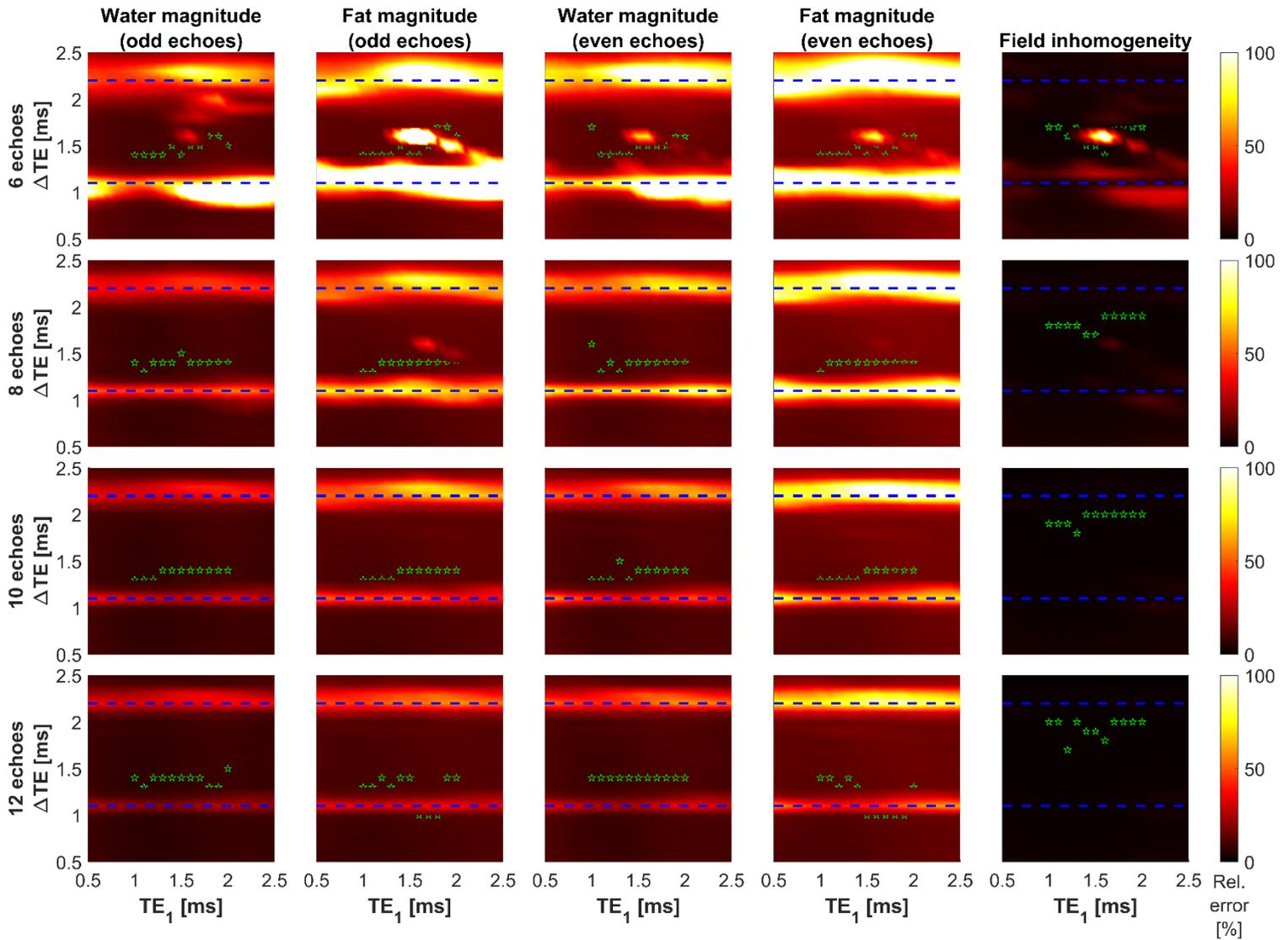

**Supplementary figure 3:** Minimum relative error heat maps as function of echo spacing and time of the first echo. Minimum was determined for calculation for PDFF 0.1, 0.5, and 0.9. Rows: Simulation for (top to bottom) 6, 8, 10, and 12 echoes. Columns: Parameters calculated for fat-water separation for odd and even echoes including magnitude of the complex fat and water components of the signal, and the field inhomogeneity term. Stars: ΔTE that generates the minimum relative error in the heat maps for each $TE_1$, estimate, and echo number. Dashed lines: range for search of optimal ΔTE (black). ΔTE=1.5 ms provides accurate estimates for 6, 8, 10, and 12-echo acquisitions.

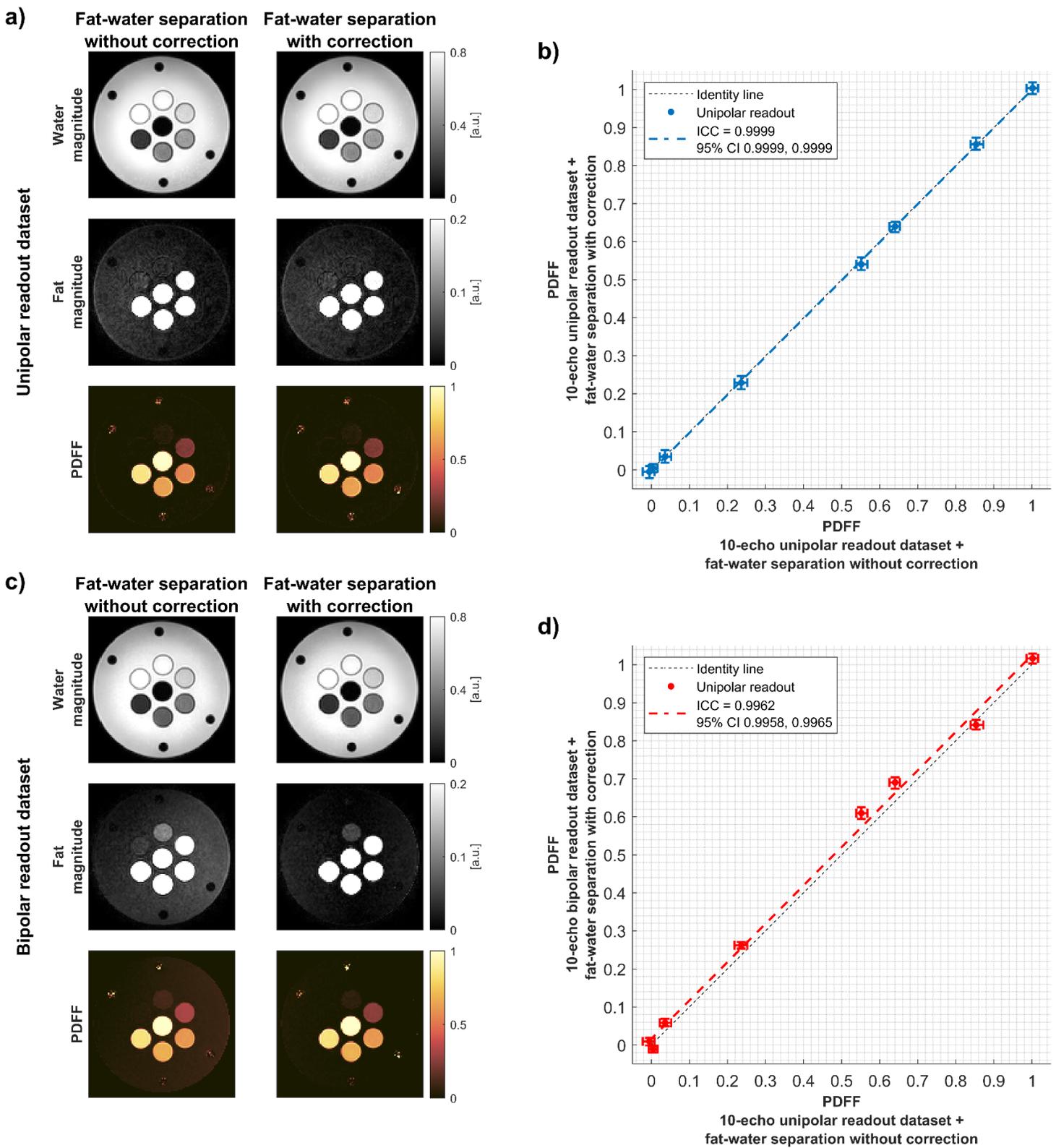

**Supplementary figure 4:** Fat-water separation performance in 6-echo data acquired with unipolar and bipolar readouts. Data processing performed with the graph-cut fat-water separation without corrections for bipolar-induced effects and with the proposed approach (with corrections). a) Results for the unipolar readout dataset. b) PDFF comparison for the unipolar dataset processed with both techniques. c) Results for the bipolar readout dataset. d) PDFF comparison for unipolar and bipolar datasets, with and without corrections, respectively. In b) and d), markers indicate the median PDFF and error bars represent the 25th–75th percentiles.

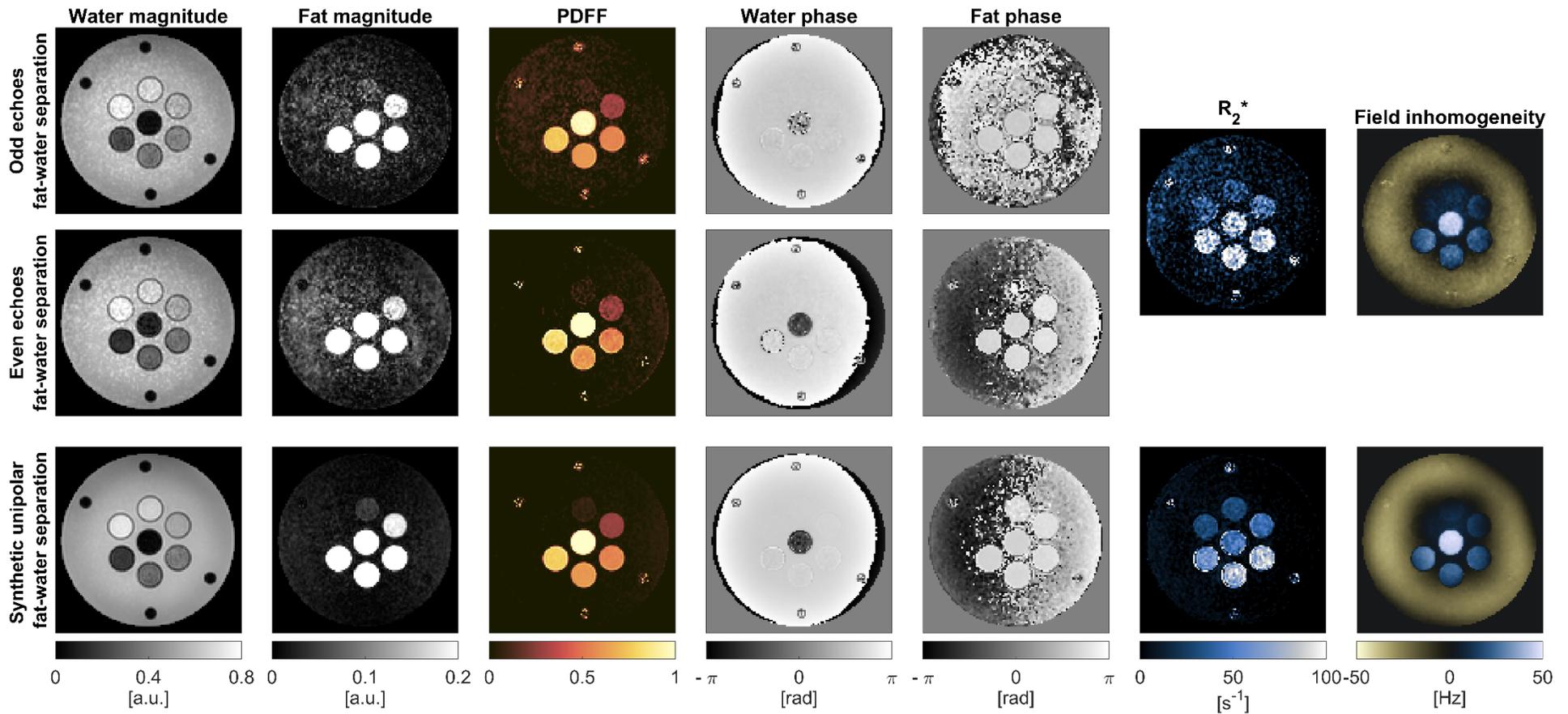

**Supplementary figure 5:** Noise and artifact reduction in fat-water separation from the synthetic unipolar dataset. From top to bottom, row present odd, even, and synthetic unipolar results for fat-water separation with a 6-echo bipolar dataset. PDFF maps (not a direct output from fat-water separation) are derived from the fat and water complex signals. Only single $R_2^*$ and field inhomogeneity maps are presented for the odd and even fat-water separation because these parameters are constrained to be the same for both cases.

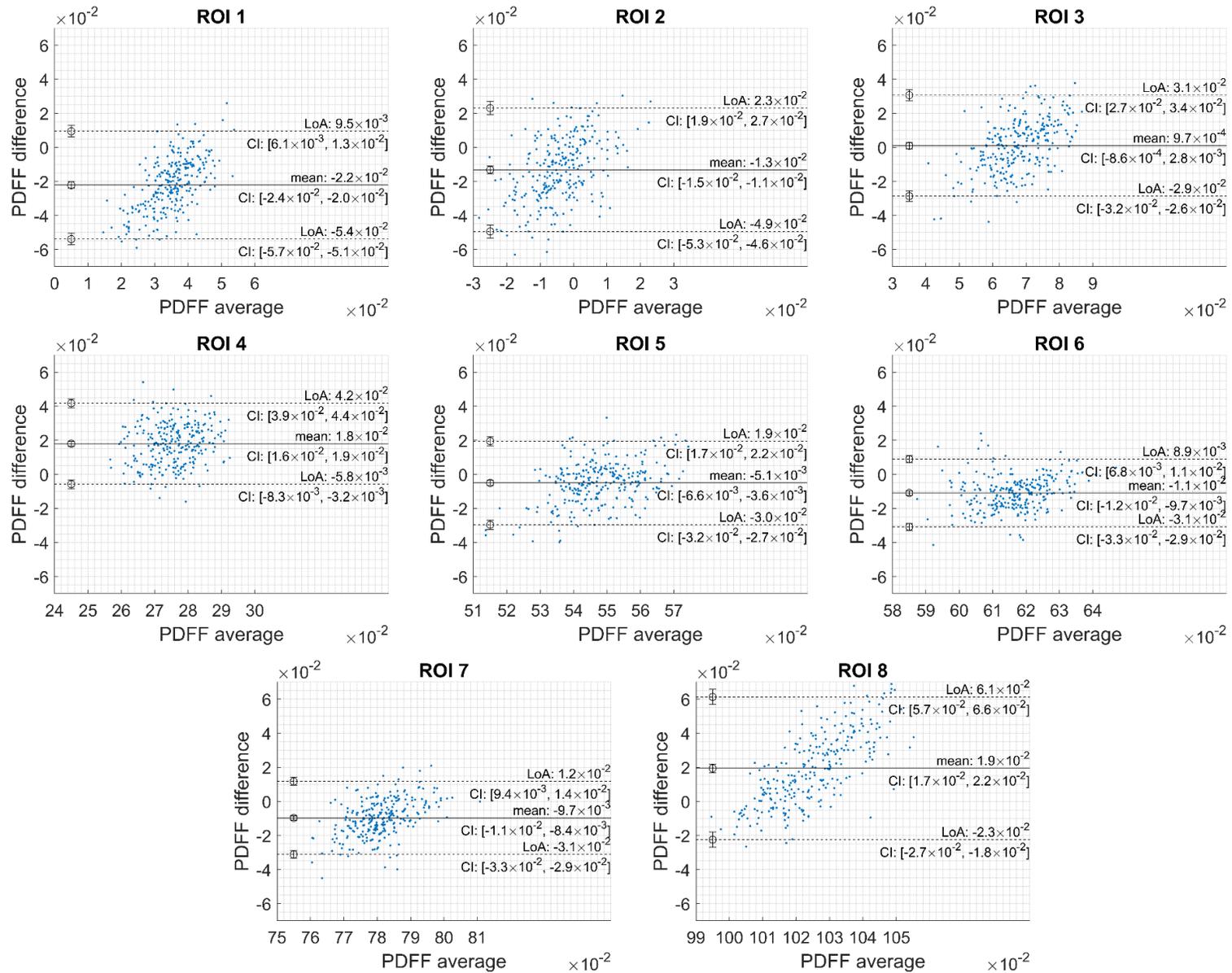

**Supplementary figure 6:** Bland-Altman plots comparing PDFF from two bipolar readout gradient datasets with 6 echoes and different echo spacing: ΔTE=0.9 ms and ΔTE=1.5 ms. Each plot compares the results in one ROI. Plots show the mean difference, limits of agreement (LoA), and their confidence intervals (CI).

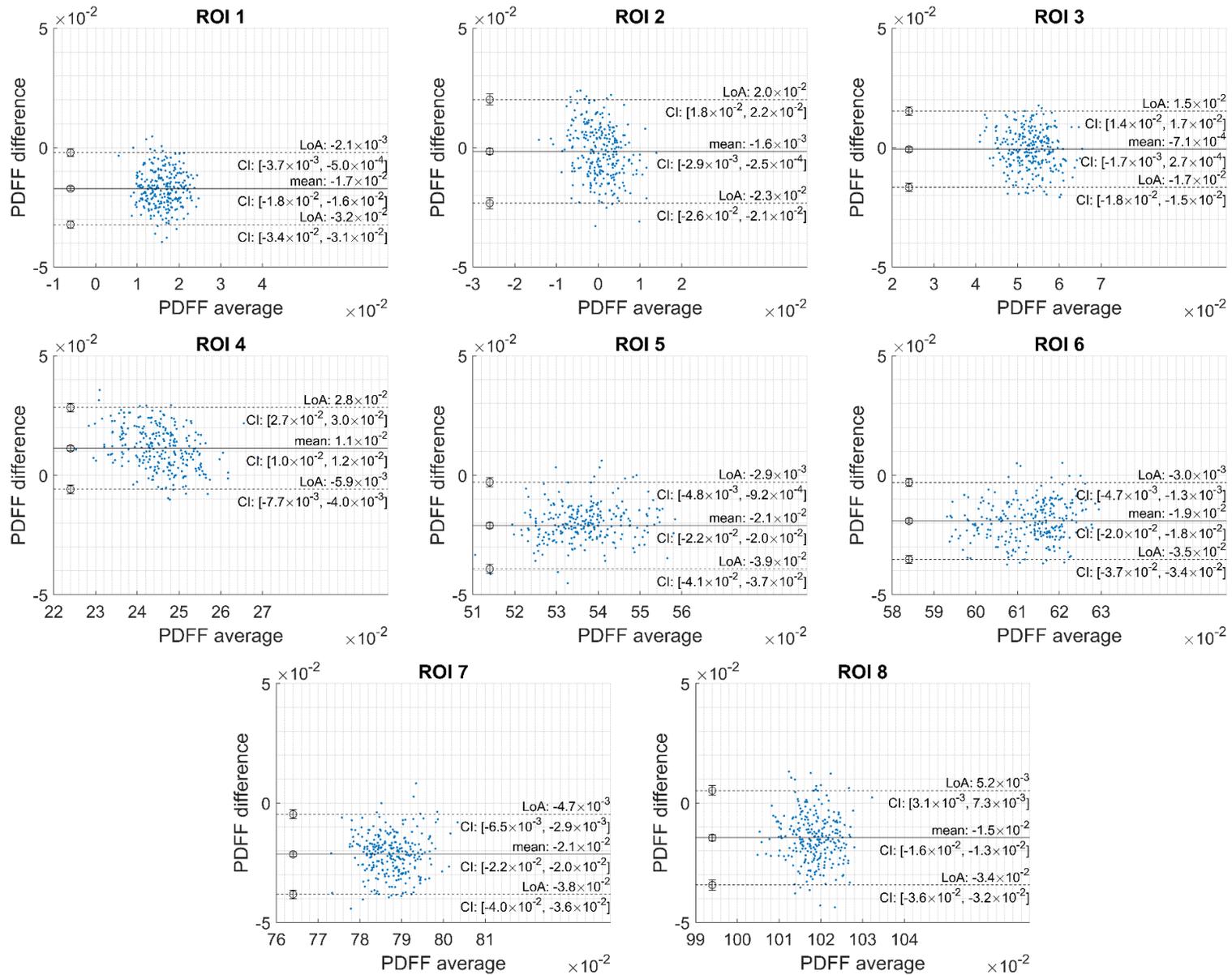

**Supplementary figure 7:** Bland-Altman plots comparing two bipolar readout gradient datasets with 10 echoes and different echo spacing: ΔTE=0.9 ms and ΔTE=1.5 ms. Each plot compares the results in one ROI. Plots show the mean difference, limits of agreement (LoA), and their confidence intervals (CI).

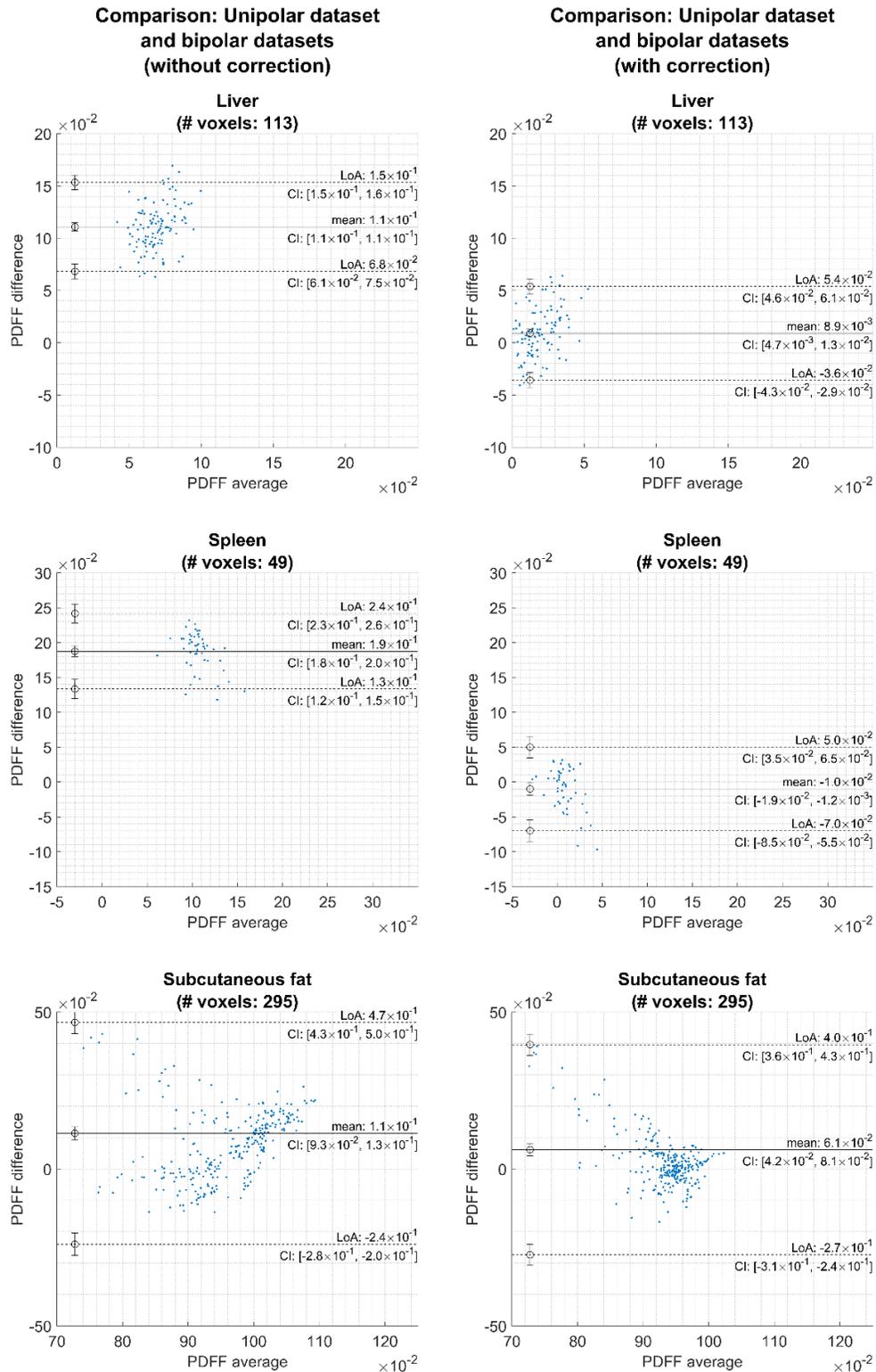

**Supplementary figure 8:** Bland-Altman plots for quantitative comparison of PDFF maps from abdominal datasets evaluated in the liver, spleen, and subcutaneous fat. Comparison considers the bipolar dataset with TR=7.2 ms and 6 echoes (experiment designed to show the application of bipolar readout gradients for reduction of TR and scan time). Plots show the mean difference, limits of agreement (LoA), and their confidence intervals (CI).

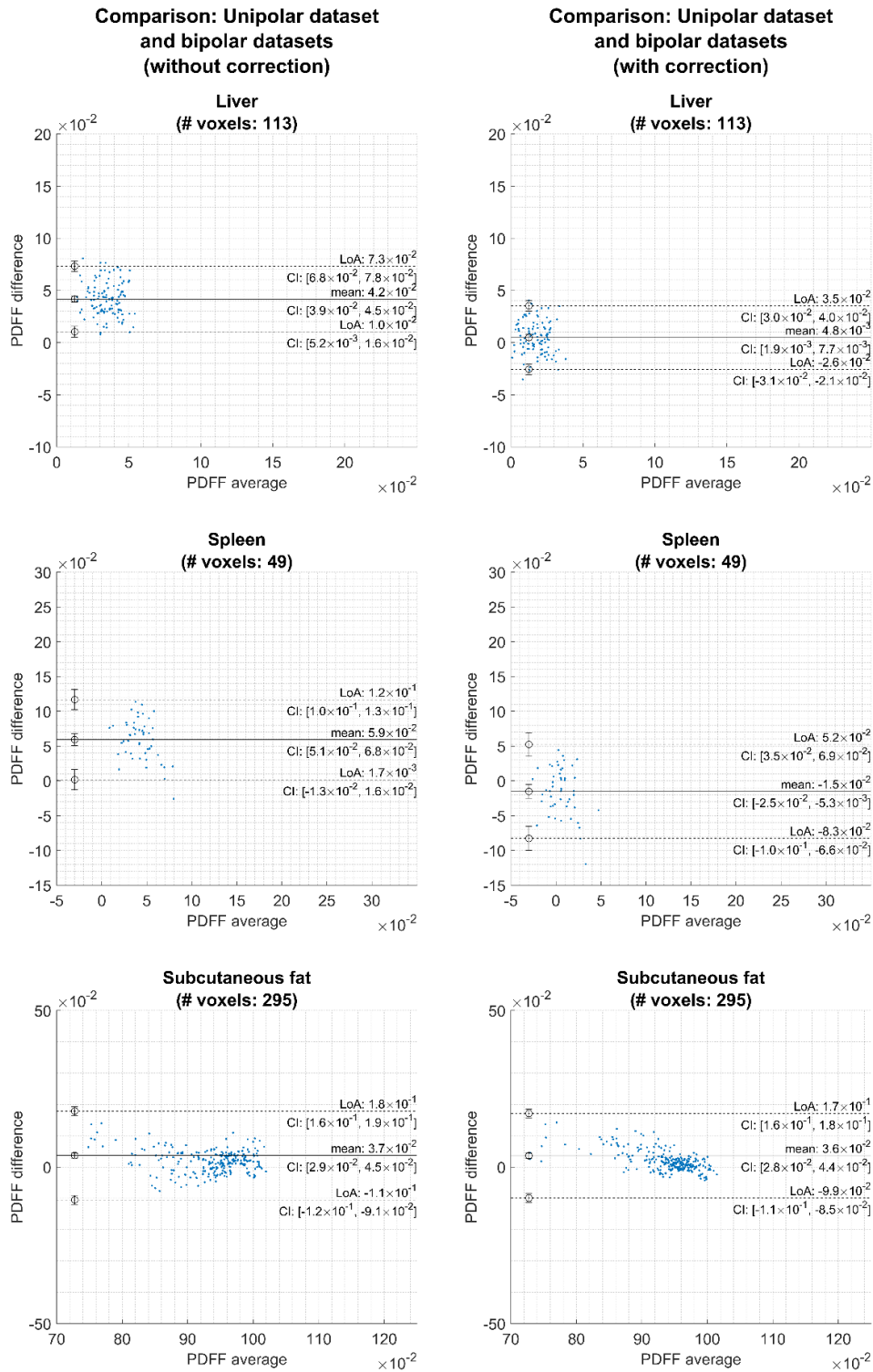

**Supplementary figure 9:** Bland-Altman plots for quantitative comparison of PDFF maps from abdominal datasets evaluated in the liver, spleen, and subcutaneous fat. Comparison considers the bipolar dataset with TR=11 ms and 10 echoes (experiment designed to show the application of bipolar readout gradients for reduction of TR and scan time). Plots show the mean difference, limits of agreement (LoA), and their confidence intervals (CI). Compared to Supplementary figure 9, the mean difference decreased for the liver and subcutaneous ROIs and LoAs decreased for all three anatomies.